\documentclass[aps,prl,twocolumn,amsmath,amssymb,superscriptaddress]{revtex4-1}
\usepackage{graphicx}
\usepackage{amssymb}
\usepackage{color}
\usepackage{amsmath}
\usepackage{float}
\usepackage{gensymb}

\usepackage{epstopdf}

\newcommand{\beq}{\begin{eqnarray}}
\newcommand{\eeq}{\end{eqnarray}}

\begin{document}

\title{Spin-charge-lattice coupling across the charge density wave transition in a Kagome lattice antiferromagnet}

\author{Xiaokun Teng}
\affiliation{Department of Physics and Astronomy, Rice University, Houston, Texas 77005, USA}

\author{David W. Tam}
\affiliation{Laboratory for Neutron Scattering and Imaging, Paul Scherrer Institut, 5232 Villigen, Switzerland}

\author{Lebing Chen}
\affiliation{Department of Physics and Astronomy, Rice University, Houston, Texas 77005, USA}

\author{Hengxin Tan}
\affiliation{Department of Condensed Matter Physics, Weizmann Institute of Science, Rehovot 7610001, Israel}

\author{Yaofeng Xie}
\affiliation{Department of Physics and Astronomy, Rice University, Houston, Texas 77005, USA}

\author{Bin Gao}
\affiliation{Department of Physics and Astronomy, Rice University, Houston, Texas 77005, USA}

\author{Garrett E. Granroth}
\affiliation{Neutron Scattering Division, Oak Ridge National Laboratory, Oak Ridge, Tennessee 37831, USA}

\author{Alexandre Ivanov}
\affiliation{Institut Laue-Langevin, 71 avenue des Martyrs CS 20156, 38042 Grenoble Cedex 9, France}

\author{Philippe Bourges}
\affiliation{Laboratoire L$\rm \acute{e}$on Brillouin, CEA-CNRS, Universit$\rm \acute{e}$ Paris-Saclay, CEA Saclay, 
91191 Gif-sur-Yvette, France}

\author{Binghai Yan}
\affiliation{Department of Condensed Matter Physics, Weizmann Institute of Science, Rehovot 7610001, Israel}

\author{Ming Yi}
\affiliation{Department of Physics and Astronomy, Rice University, Houston, Texas 77005, USA}

\author{Pengcheng Dai}
\email{pdai@rice.edu}
\affiliation{Department of Physics and Astronomy, Rice University, Houston, Texas 77005, USA}

\begin{abstract}
Understanding spin and lattice excitations in a metallic magnetic ordered system form the basis to unveil the magnetic and lattice exchange couplings and their interactions with itinerant electrons. Kagome lattice antiferromagnet FeGe is interesting because it displays rare charge density wave (CDW) deep inside the antiferromagnetic ordered phase that interacts 
with the magnetic order. We use neutron scattering to study the evolution of spin and lattice excitations across the CDW transition $T_{\rm CDW}$ in FeGe. While spin excitations below $\sim$100 meV can be well described by
spin waves of a spin-1 Heisenberg Hamiltonian, spin excitations at higher energies are 
centered around the Brillouin zone boundary and extend up to $\sim180$ meV consistent with quasiparticle excitations across 
spin-polarized electron-hole Fermi surfaces. Furthermore, $c$-axis spin wave dispersion and Fe-Ge optical phonon modes show a clear 
hardening below $T_{\rm CDW}$ due to spin-charge-lattice coupling but with no evidence for a phonon Kohn anomaly. By comparing our experimental results 
 with density functional theory calculations in absolute units, we conclude that FeGe is a Hund's metal
in the intermediate correlated regime where magnetism has contributions from both itinerant and localized electrons arising from spin polarized electronic bands near the Fermi level. 
\end{abstract}

\maketitle

In insulating magnets where unpaired electrons are localized on magnetic atomic sites, interactions of local spin moments 
are governed by the Heisenberg exchange couplings \cite{heisenberg}. Magnons arising from spin vibrations about their equilibrium 
positions should be characterized by linearized spin wave theories that ignore all terms of order higher than quadratic 
and interactions with lattice vibrations \cite{PhysRev.58.1098,PhysRev.117.117}. For a magnetic ordered material with more than
one magnetic ion per unit cell, we expect to observe acoustic and optical spin waves, just like
acoustic and optical phonon modes are expected for a crystalline solid with more than one atom per unit cell \cite{Boothroyd}.For example, spin waves in insulating kagome \cite{PhysRevLett.115.147201} and honeycomb \cite{PhysRevX.8.041028} lattice ferromagnet 
have well-defined acoustic and optical modes as expected for a local moment Heisenberg magnet. A spin gap between the acoustic and optical spin waves at the Dirac points can give rise
to protected topological magnon bands and edge modes \cite{PhysRevLett.115.147201,PhysRevX.8.041028}.

For metallic magnets, magnetic order can arise from either localized moments similar to an insulating magnet 
or quasiparticle spin-flip excitations between the valence (hole) and conduction (electron) bands at the Fermi level, dubbed a spin density
wave (SDW), due to electron-electron correlations \cite{RevModPhys.66.1}. Spin excitations from these magnets can arise from 
vibrations of localized moments \cite{heisenberg} or electron-hole Fermi surface quasiparticle excitations of itinerant electrons
 \cite{Bloch,PhysRev.49.537,Stoner}, respectively. Since SDW order  
can coexist and intertwine with other orders such as charge density wave (CDW) \cite{RevModPhys.60.209} and superconductivity 
\cite{RevModPhys.87.457,RevModPhys.87.855}, a determination of the interplay between magnetic and other intertwined orders forms the
basis to understand correlated electron materials. 

Recently, kagome lattice magnet FeGe was found to have a CDW order deep inside the magnetic ordered phase that couples with the ordered moment (Fig. 1a) \cite{doi:10.1143/JPSJ.18.589,bernhard_neutron_1984,bernhard_magnetic_1988,teng_discovery_2022,PhysRevLett.129.166401,teng_magnetism_2023}.
 With decreasing temperature, FeGe first exhibits collinear $A$-type antiferromagnetic (AFM) order at $T_{\rm N}\approx 400$ K (Fig. 1b), forms a $2\times 2\times 2$ CDW order below $T_{\rm CDW}\approx 110$ K with an enhanced ordered moment, and finally develops incommensurate AFM structure below $T_{\rm Canting} \approx 60$ K (Fig. 1c) 
\cite{bernhard_neutron_1984,bernhard_magnetic_1988,teng_discovery_2022,PhysRevLett.129.166401,teng_magnetism_2023,miao_spin-phonon_2022}. In previous angle resolved photoemission spectroscopy (ARPES) and inelastic neutron scattering experiments,
electron-boson interaction induced kink around 30 meV seen in ARPES spectra was identified as electron and optical phonon coupling \cite{teng_magnetism_2023}. From temperature dependent 
low-energy spin wave measurements, incommensurate spin fluctuations associated with incommensurate AFM 
static order show a kink at $T_{\rm CDW}$ and survive up to $T_{\rm N}$ \cite{chen_competing_2023}. Although these results suggest that incommensurate AFM order is actually a SDW phase instead of the double cone AFM structure
as suggested originally \cite{bernhard_neutron_1984,bernhard_magnetic_1988}, there is no determination of 
the spin-charge-lattice coupling across $T_{\rm CDW}$ \cite{teng_magnetism_2023} and the 
microscopic origin of the magnetic order. Using first principle calculations, it was predicted that
the nearest-neighbor magnetic exchange interactions are dominate and ferromagnetic (FM), incommensurate 
AFM order is due to lattice distortion-induced Dzyaloshinskii-Moriya (DM) interactions \cite{PhysRevB.108.035138}.
Furthermore, spin waves should be strongly dispersive in the kagome plane with acoustic mode below about 100 meV 
and two optical modes extending up to 250 meV \cite{PhysRevB.108.035138}, similar to optical modes in 
insulating kagome lattice magnet \cite{PhysRevLett.115.147201}.

Here we use inelastic neutron scattering to study the evolution of spin and lattice excitations of FeGe across $T_{\rm CDW}$ \cite{SI}. While spin excitations below $\sim$100 meV can be well described by a spin-1 ($S\approx 1$) local moment Heisenberg Hamiltonian, spin excitations at higher energies are centered around the Brillouin zone boundary and extend up to $\sim180$ meV (Figs. 1d-f), clearly different from the predictions of the first principle calculations \cite{PhysRevB.108.035138} and the local moment picture. Instead, the high energy spin waves are consistent with quasiparticle excitations between spin-polarized electron-hole Fermi surfaces, similar to spin excitations in metallic antiferromagnets FeSn \cite{xie_spin_2021,PhysRevB.105.L180403,PhysRevB.107.174407} and Fe$_{0.89}$Co$_{0.11}$Sn ~\cite{PhysRevB.106.214436}, van der Waals metallic ferromagnet Fe$_{2.72}$GeTe$_2$ \cite{bao_neutron_2022}, weak itinerant ferromagnet MnSi \cite{chen_unconventional_2020,doi:10.1126/sciadv.add5239}, and heavy Fermion CePd$_3$ \cite{science_aan0593}. Furthermore, spin wave dispersions and Fe-Ge optical phonon modes show a clear hardening below $T_{\rm CDW}$ due to spin-charge-lattice coupling but with no evidence for phonon Kohn anomaly. By comparing these results with density functional theory (DFT) calculations in absolute units, we conclude that FeGe has an intimate coupling between itinerant electrons and magnetism similar to iron pnictides \cite{Dai_2012}, suggesting that it is a Hund's metal with intermediate electronic correlations \cite{doi:10.1146/annurev-conmatphys-020911-125045,STADLER2019365}. The strong spin-charge-lattice coupling in FeGe is different from FeSn \cite{xie_spin_2021,PhysRevB.105.L180403,PhysRevB.107.174407} and all other kagome lattice materials \cite{WOS:000934065100013}, making this an unique system to investigate intertwined orders in spin, charge, and lattice degrees of freedom.

\begin{figure}[ht]
	\centering
	\includegraphics[scale = 0.85]{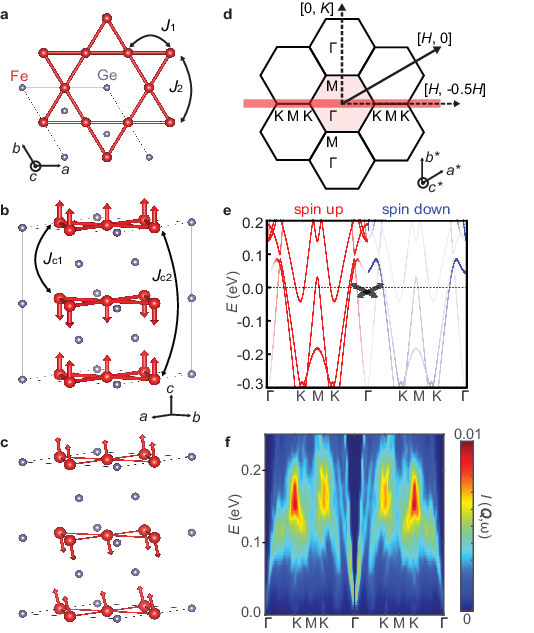}
	\caption{(a) In-plane lattice structure of FeGe with red and grey denoting Fe and Ge atoms respectively. Magnetic structure of FeGe at (b) 
	$T > T_{\rm Canting}$ and (c) $T < T_{\rm Canting}$. (d) Reciprocal space of FeGe. The thick red line denotes the momentum path 
	along the $[H, -0.5H, -1.5]$ direction. (e) DFT (with no spin orbit coupling) calculation. The red and blue bands represent spin up and down bands respectively. Thick arrows indicate particle-hole (\textit{p}-\textit{h}) scattering. (f) Dynamic susceptibility calculated from \textit{p}-\textit{h} scattering of the set of bands in (e), where the energy axis has been renormalized by 1.7 \cite{teng_magnetism_2023}. } 
	\label{fig:Figure1}
 \end{figure}

We first examine energy ($E$) and momentum ($\bf Q$) dispersion of the in-plane spin excitations of FeGe in the $A$-type AFM ordered phase at 120 K. At low energy, spin-wave-like excitations stem from the $\Gamma$ point at the zone center and gradually disperse to the zone boundary, first reaching the $M$ points at $E = 90\pm10$ meV and then the $K$ points at $E = 120\pm 10$ meV, showing intensity modulation across zone boundaries (Figs. 1d and 2a). Figure 2b shows the spin excitation dispersion along the high-symmetry direction $[H, -0.5H, -1.5]$ with $L = [-3, 0]$ as defined in Fig. 1d taken with incident neutron energy of $E_i = 300$ meV. We take systematic constant energy cuts from $E_i = 45$, 100, and 300 meV data at temperatures across $T_{\rm CDW}$ and $T_{\rm Canting}$, and fit them with two Gaussians with a linear background. After averaging between the left and right, we obtain the spin wave dispersions at different temperatures, and find no significant change change across $T_{\rm CDW}$ and $T_{\rm Canting}$ from 120 K to 8 K (Fig. 2c).

\begin{figure}[ht] 
	\includegraphics[scale = 1]{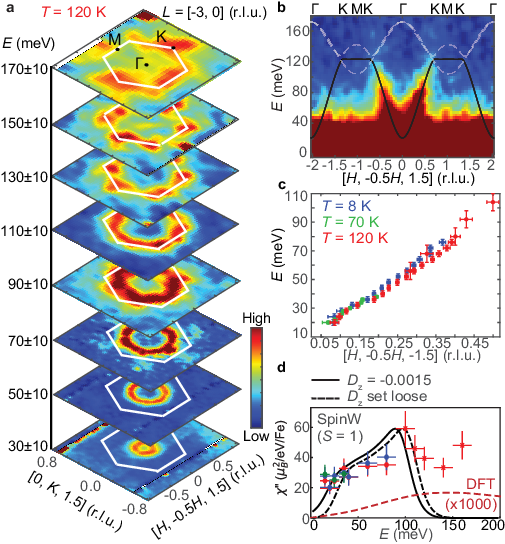}
	\centering
	\caption{(a) Constant energy slices at $E=30\pm 10$, $50\pm 10$, $70\pm 10$, $90\pm 10$, $110\pm 10$, $130\pm 10$, $150\pm 10$, $170\pm 10$ meV. The white lines are Brillouin zone boundaries, and high symmetry points are labeled with black points. (b) In-plane magnetic excitation 
	along the $[H, -0.5H, -1.5]$ direction. The black solid lines and gray dashed lines indicate simulated acoustic and optical spin wave branches, respectively, using a Heisenberg model (SpinW) \cite{toth_linear_2015}. Data in (a) and (b) is taken at 120 K. (c) Spin wave dispersions extracted by Gaussian fitting with a linear background of the symmetrized spectra along the $[H, -0.5H, -1.5]$ direction. The vertical error bars indicate the energy integration range. Horizontal error bars are from fitting. Integration along the orthogonal in-plane direction is [-0.1, 0.1]. (d) Integrated magnetic intensity in the first in-plane Brillouin zone and averaged between $L = [-2.5, -0.5]$. The black line indicates the calculated spin wave intensity in absolute units assuming $S$ = 1 in the SpinW + Horace program \cite{toth_linear_2015}. Red, green and blue symbols represent 120 K, 70 K and 8 K data respectively \cite{EWINGS2016132}. The red-dashed line is the DFT estimated 
	spin susceptibility from the two electronic bands in Figs. 1f and 1e.}
	\label{fig:Figure2}
\end{figure}

To understand magnetic excitations using a local moment picture, we consider the Heisenberg Hamiltonian:
\[H_{0} = \sum_{<i,j>} J_{ij}S_{i}\cdot S_{j} + \sum_{i} D_{z}(S_{z_{i}})^2\]
Here, $J_{ij}$ represents the magnetic exchange interaction between the $i$th and $j$th Fe atoms, $S_{i}$ and $S_{j}$ are the local spins at the $i$th and $j$th sites, respectively, and $D_z$ is the single-ion anisotropy. For in-plane dispersions, we use $J_1$ and $J_2$ as the nearest and next-nearest-neighbor couplings, respectively (Fig. 1a). The $c$-axis nearest and next-nearest-neighbor couplings $J_{c1}$ and $J_{c2}$ (Fig. 1b) are determined in Ref. \cite{chen_competing_2023}. We can simulate spin wave dispersions using the above Heisenberg Hamiltonian and compare with experiments.

\begin{table}[]
    \centering
    \begin{tabular}{|c|c|c|c|c|c|}
    \hline
       Model &  $J_{1}$ &  $J_{2}$ &  $J_{c1}$ & $J_{c2}$ & $D_z$ (meV)  \\
       \hline
        Heisenberg &  $-16.4$ &  $-7.2$ &  $11.3$  & 0& $-0.015$ \\
        \hline
        First principle &  $-41.97$ &  $5.49$ &  $8.44$  & $-2.04$ & 0 \\
        \hline
    \end{tabular}
    \caption{The first row gives magnetic exchange coupling constants obtained from the Heisenberg Hamiltonian simulation. The solid and dashed lines in Fig. 2b shows the resulting spin wave dispersions. The second row shows exchange parameters predicted from the first principle calculations \cite{PhysRevB.108.035138}.}
    \label{tab:my_label}
\end{table}

In previous low-energy inelastic neutron scattering experiments, the dispersion of spin excitations along the $L$ direction was found to have a band top around $E\approx 24$ meV and a small $D_z\approx -0.015$ meV \cite{chen_competing_2023}. To better fit the high-energy in-plane
spin excitation data, the anisotropic term was set loose in the fitting process due to the large $L$ integration range with $E_i = 300$ meV. The solid black lines in Fig. 2b represent the acoustic spin-wave branches that are in agreement with the data below $\sim$100 meV with magnetic exchange couplings in Table 1. However, two distinctions between the data and the simulation cannot be explained by the Heisenberg model. First, the optical branches predicted by the Heisenberg model and first principle calculations \cite{PhysRevB.108.035138} are absent in the data (Fig. 2b). Second, when approaching the Brillouin zone boundary at $K$ (Dirac points) and $M$ points (Figs. 2b, 2c), the magnetic excitations continue to rise and form a convex shape, instead of bending over to a concave shape as predicted by the simulation. As a consequence, 
there is no spin gap at the Dirac points and we do not expect spin excitations of FeGe to have 
protected topological magnon bands and edge modes \cite{PhysRevLett.115.147201,PhysRevX.8.041028}. Figure 2d compares the integrated local dynamic spin susceptibility in the first Brillouin zone at 8 K, 70 K, and 120 K with $L = [0.5,2.5]$, normalized by acoustic phonons near a nuclear Bragg peak to absolute unit \cite{xu_absolute_2013}, with calculated results from Heisenberg model assuming $S = 1$. The susceptibility shows a broad peak around 100 meV, and decays rapidly for energies above 100 meV (Fig. 2d). Although DFT calculations of the hole-electron quasiparticle excitations (Figs. 1e and 1f) correctly predicted the high energy spin excitations, the absolute magnetic scattering estimated from the DFT 
is much smaller than the observation (Fig. 2d), similar to iron pnictides \cite{WOS:000329395700003}. 

\begin{figure}[ht]
	\centering
	\includegraphics[scale = 1]{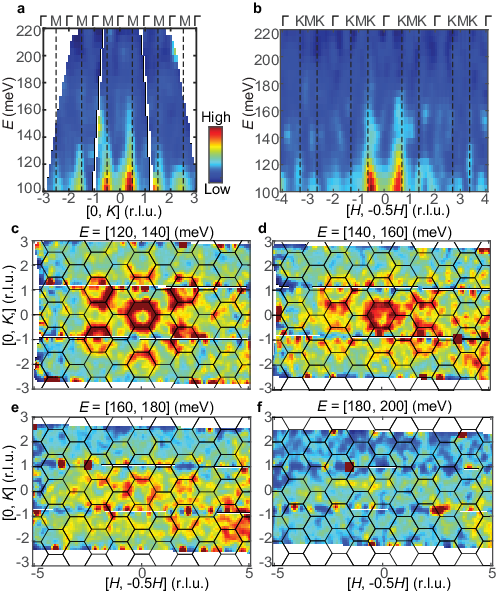}
	\caption{High energy magnetic excitations along (a) $[0, K]$ with $[H, -0.5H] = [-0.1, 0.1]$ and (b) $[H, -0.5H]$ 
	with $K = [-0.1, 0.1]$, $L = [-4, 0]$ (r.l.u.) measured with $E_i = 300$ meV. The dashed lines indicate the $M$ points in (a) and $K$ points in (b). Constant energy slices at (c) $E=130\pm 10$, (d) $150\pm 10$, (e) $170\pm 10$ meV and (f) $190\pm 10$ meV. The solid black lines mark the Brillouin zone boundaries. Data is taken at 120 K.}
	\label{fig:Figure3}
\end{figure}

Since spin excitations above 120 meV in FeGe are rod-like at both the $M$ (Fig. 3a) and the $K$ points (Fig. 3b), they are similar to metallic antiferromagnets FeSn ~\cite{xie_spin_2021,PhysRevB.105.L180403,PhysRevB.107.174407,PhysRevB.106.214436}, ferromagnets Fe$_{2.72}$GeTe$_2$ \cite{bao_neutron_2022}, MnSi \cite{chen_unconventional_2020,doi:10.1126/sciadv.add5239}, and heavy Fermion CePd$_3$ \cite{science_aan0593}.
To further investigate the behavior of this rod-like dispersion, we plot the constant energy slices in multiple Brillouin zones (Figs. 3c-f). The magnetic excitations are concentrated at Brillouin zone boundary (Figs. 3c-e), clearly different from that of MnSi where the rod-like dispersions reside inside Brillouin zone boundaries \cite{chen_unconventional_2020,doi:10.1126/sciadv.add5239}. 
Spin excitations gradually vanish around $E$ = 200 meV (Fig. 3f),
different from expectations of a local moment Heisenberg Hamiltonian and the first principle calculations \cite{PhysRevB.108.035138}. These results are similar to high-energy spin excitations of FeSn, which also has $A$-type AFM but without CDW order~\cite{xie_spin_2021,PhysRevB.105.L180403}. The wave vector dependence of these excitations are clearly different from cluster spin excitations associated with localized
spins in insulating frustrated pyrochlore \cite{Lee_2002} and triangular lattice antiferromagnets 
\cite{Gao_2023}, as well as in high energy spin excitations in metallic kagome lattice ferrimagnet 
TbMn$_6$Sn$_6$ \cite{riberolles_low_2022,riberolles_chiral_2023}.

\begin{figure}[ht]
	\centering
	\includegraphics[scale = 1]{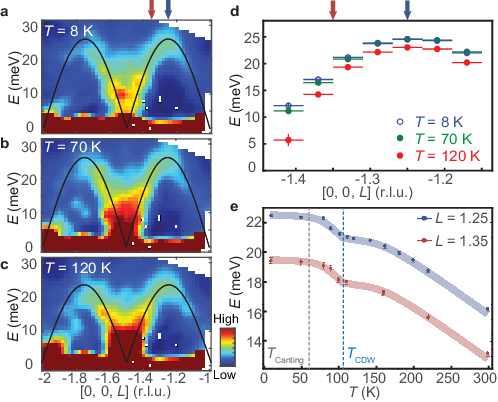}
	\caption{Temperature dependent spin excitations along the $c$-axis ($L$ direction) at (a) 8 K, (b) 70 K and (c) 120 K measured 
	with $E_i = 45$ meV. (d) Spin wave dispersions determined by fitting constant momentum cuts taken from (a-c) with 
Lorentzian on a constant background.
Red and blue arrows in (a) and (d) mark $L = 1.35$ and $L = 1.25$, respectively.
 (e) Temperature dependent spin wave energy at $L = 1.35$ (blue dots) and 1.25 (red dots), obtained from IN8. 
Grey and blue dashed lines mark $T_{\rm Canting}$ and $T_{\rm CDW}$ respectively. Thick blue and red lines are guides to the eye.}
	\label{fig:Figure4}
\end{figure}

Figures 4a-c show the out-of-plane spin wave excitations at different temperatures across $T_{\rm CDW}$ and $T_{\rm Canting}$. The fitted spin wave dispersions are shown in Fig. 4d. 
By extracting the band top $E_{top}$ at $L = -1.25$ for each temperature,
we find that spin waves harden by around 11.6\% from 120 K to 70 K across $T_{\rm CDW}$, but remain unchanged across $T_{\rm Canting}$. The temperature dependence of the
spin wave energy measured at $L = 1.25$ and $L = 1.35$ near the zone boundary 
using a triple-axis spectrometer 
shows a clear hardening of spin wave dispersion around 2 meV below $T_{\rm CDW}$ (Fig. 4e).
 This hardening of spin waves coincides with an increase of around 0.1 $\mu_B$/Fe in the static magnetic moment, from around 1.5 $\mu_B$/Fe to 1.6 $\mu_B$/Fe across $T_{\rm CDW}$ \cite{teng_discovery_2022}. The behavior is not as prominent for the in-plane dispersion because of its steeper dispersion \cite{SI}. Therefore, the hardening of spin wave dispersion below $T_{\rm CDW}$ 
arises from the CDW order induced
moment increase and spin waves from the $A$-type AFM order mostly conform
to a local moment Hamiltonian below about 100 meV.
 
\begin{figure}[ht]
	\centering
	\includegraphics[scale = 1]{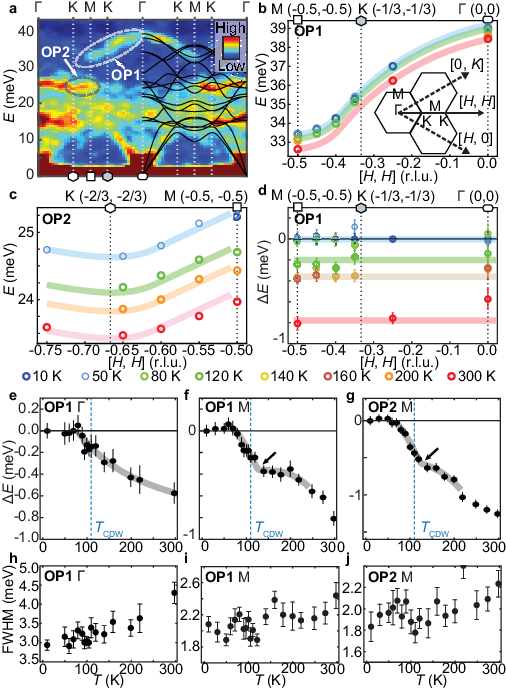}
	\caption{(a) In-plane phonon dispersion at $L = [-3.1, -2.9]$ and 120 K.
	The solid black lines are DFT calculations.	Two separate optical phonon 
	branches are labeled as OP1 and OP2, corresponding to Fe-Ge $A_{2u}$ and Fe out-of-plane vibrations, respectively. Phonon dispersions of (b) the OP1 and (c) OP2 at 50 K, 120 K, 200 K and 300 K. 
	(d) Temperature dependence of the OP1 energy shift throughout the Brillouin zone at 10 K, 50 K, 80 K,
	120 K, 140 K, 160 K, and 300 K.
	The phonon energy at the base temperature (10 K) is subtracted. 
Temperature dependence of the OP1 phonon mode at (e) $\Gamma$, (f) $M$ points, and the OP2 phonon mode at $M$ point.
The thick colored lines in (b-d) and the gray lines in (e-g) are guides to the eye. 
Temperature dependence of phonon full width half maximum (FWHM) at the OP1 $\Gamma$ (h), OP1 $M$ (i) and OP2 $M$ 
	(j). }
	\label{fig:Figure5}
\end{figure}

After mapping out spin excitation evolution across $T_{\rm CDW}$ and $T_{\rm Canting}$, 
we investigate temperature dependence of the in-plane phonon spectra (Figs. 5a-5j) \cite{SI}. In previous 
Raman and neutron Larmor diffraction measurements, the crystal structure is found to change
from hexagonal to monoclinic with a small in-plane lattice distortion, but becomes more symmetric below $T_{\rm CDW}$ \cite{wu2023symmetry}.
However, much is unclear on the dynamic spin-lattice-charge interactions acrossing the CDW transition. 
Figure 5a shows the overall phonon spectra where optical phonon 1 (OP1) and optical phonon 2 (OP2) are marked.
They agree well with the DFT calculated spectra shown in solid black lines. 
For both OP1 and OP2, the entire phonon dispersion hardens about 1 meV at all measured 
wave vectors on cooling from 300 K to 10 K (Figs. 5b-5d), distinct from previous work on acoustic phonon mode 
where no phonon energy shift is observed at $M$ points \cite{miao_spin-phonon_2022}. To further determine if the phonon hardening is coupled with
CDW order, we plot the temperature dependence of phonon energy at high symmetry $\Gamma$ and $M$ points for OP1 and
OP2 (Figs. 5e-5g). There is almost no shift in phonon dispersion below around 80 K, and CDW order 
is clearly coupled with phonon hardening at $M$ points for OP1 and OP2 modes (Figs. 5f and 5g). 
Since OP1 is the optical $A_{2u}$ mode involving out of plane Fe-Ge vibrational modes and OP2 is Fe out of plane vibrational mode,
the results suggest a strong coupling of these modes with the CDW order associated the Ge $c$-axis dimerization \cite{miao_spin-phonon_2022}.
 Note that the phonon full width half maximum (FWHM) at different wave vectors consistently decreases with decreasing temperature (Fig. 5h-j), corresponding to a larger phonon lifetime at lower temperatures. For comparison, acoustic zone boundary 
phonon lifetime was found to decrease at $A$ point from 400 K to 200 K but not at $M$ point
from inelastic X-ray scattering experiments \cite{miao_spin-phonon_2022}. 

In previous work on FeGe \cite{miao_spin-phonon_2022}, 
the CDW transition is suggested to arise from Ge $c$-axis dimerization different from
the usual Kohn anomaly in electron-phonon coupled CDW materials but similar to spin-lattice coupling in 
FeSi \cite{WOS:000366381200001} and CuGeO$_3$ \cite{PhysRevLett.80.3634,PhysRevB.58.R14677}. 
These results are consistent with first principle calculations \cite{PhysRevMaterials.7.104006,doi:10.1021/acsnano.3c00229,zhang2023triplewell} and subsequent experiments \cite{chen2023longranged,zhao2023photoemission}, suggesting that CDW order in FeGe arises from electron correlations instead of the usual electron-phonon interaction \cite{Wu:1171103}. 
From our inelastic neutron scattering experiments summarized in Figs. 2-5, we find that high energy spin excitations of FeGe behave similarly to other itinerant magnets \cite{xie_spin_2021,PhysRevB.105.L180403,PhysRevB.107.174407,PhysRevB.106.214436,bao_neutron_2022,chen_unconventional_2020,doi:10.1126/sciadv.add5239,science_aan0593}.  The lack of optical spin waves and rod-like 
spin excitations confined to the Brillouin zone boundary for energies above 
100 meV (Fig. 3) are consistent with quasiparticle excitations from 
spin down to spin up bands across the Fermi level as shown in our 
DFT calculations (Figs. 1e and 1f), 
indicating that itinerant electrons play an important role in determining 
the high energy spin excitations. However, the magnitude of the magnetic scattering determined from DFT is much smaller 
than that of the observation (Figs. 1f and 2d). For comparison, we note that high energy spin excitations
in iron pnictides are consistent with a local moment picture while low-energy spin excitations 
are from the nesting of electron and hole Fermi surfaces and DFT calculated spectral weight is also much 
smaller than the observation \cite{Dai_2012,WOS:000329395700003}. Since the DFT calculated electronic dispersions need to be renormalized by about 1.7 to account for those determined from ARPES experiments \cite{teng_magnetism_2023} similar to the values in iron pinctide superconductors \cite{WOS:000297692900017}, we conclude that 
FeGe is a Hund's metal in the intermediate 
electron correlation regime. Similarly, while 
a pure local moment Heisenberg model with $S=1$ can account for temperature dependent spin 
excitations above the anisotropy gap energy of $\sim$1 meV \cite{chen_competing_2023}, spin excitations above 100 meV appear to have an itinerant origin (Fig. 2d). These results are consistent with recent theoretical calculations indicating that FeGe is slightly more electron correlated
compared with FeSn \cite{lin2023complex}, and the flatish electronic bands near the Fermi level responsible for correlated properties of FeGe
arise from $(d_{xy}, d_{x^2-y^2})$ and $(d_{xz},d_{yz})$ orbitals \cite{jiang2023kagome}. 
Therefore, FeGe is in the intermediate correlated regime where magnetism has contributions from both itinerant and localized electrons, and couples with CDW order to form a strong spin-charge-lattice coupled kagome metal, suggesting that FeGe is a rare case where the energy scales of spin, charge, lattice degrees of freedom are similar and their interactions give rise to the observed exotic properties 
different from its sister compound FeSn and all other known kagome lattice magnets \cite{WOS:000934065100013}.
 
We are grateful to Andrea Piovano for helpful discussions. The neutron scattering and single crystal synthesis work at Rice was supported by US NSF DMR-2100741 and by the Robert A. Welch Foundation
under grant no. C-1839, respectively (P.D.). M.Y. acknowledges
support by the U.S. DOE grant No. DESC0021421
and the Robert A. Welch Foundation, Grant
No. C-2175. A portion of this research used resources
at the Spallation Neutron Source, a DOE Office of Science
User Facility operated by Oak Ridge National Laboratory.

%It is also worth noting that the FWHM does not show any trend across different transitions \cite{SI}, indicating that there is no significant change in the lifetime of the $L$ magnons.

%%Bibliography
\bibliographystyle{apsrev4-1}
\bibliography{FeGe_INS}

\end{document}